\documentclass[aip,jcp,amsmath,amssymb,citeautoscript,preprint]{revtex4-1}

\usepackage{dcolumn}
\usepackage{bm}
\usepackage{graphicx}
\usepackage{longtable}
\usepackage{footnote}
\usepackage{float}

\usepackage{color}

\draft 

\begin{document}
\raggedbottom

\newcommand{\bra}[1]{\langle #1|}
\newcommand{\ket}[1]{|#1\rangle}
\newcommand{\braket}[2]{\langle #1|#2\rangle}
\newcommand{\p}{^\prime}
\newcommand{\pp}{^{\prime\prime}}

\newcommand{\etal}{{\it et al.} }
\newcommand{\ai}{{\it ab initio}}
\newcommand{\vi}{{\it vide infra}}
\newcommand{\cm}{cm$^{-1}$}
\newcommand{\BODC}{DBOC}
\newcommand{\ownPES}{XBRD}
\newcommand{\Hartree}{{\it E}$_{\rm h}$}
\newcommand{\hp}{H$_3^+$}

\def\dboc{{DBOC}}
\def\a0{{$a_{\rm 0}$}}

\newcommand{\Cv}[1]{${\mathcal C}_{#1{\rm v}}$}
\newcommand{\Dh}[1]{${\mathcal D}_{#1{\rm h}}$}
\newcommand{\Ch}[1]{${\mathcal C}_{#1{\rm h}}$}
\newcommand{\Dd}[1]{${\mathcal D}_{#1{\rm d}}$}

\newcommand{\Cvv}{${\mathcal C}_{2{\rm v}}$}
\newcommand{\Cvvv}{${\mathcal C}_{3{\rm v}}$}
\newcommand{\Cs}{${\mathcal C}_{\rm s}$}
\newcommand{\lnr}{^{\ell}}
\newcommand{\Dhhh}{${\mathcal D}_{3{\rm h}}$}
\newcommand{\Td}{${\mathcal T}_{\rm d}$}
\newcommand{\Oh}{${\mathcal O}_{\rm h}$}

\newcommand{\2}{$_{2}$}
\newcommand{\3}{$_{3}$}
\newcommand{\4}{$_{4}$}

\newcommand{\hpph}{P$_2$H$_2$}
\newcommand{\trans}{\textit{trans}}
\newcommand{\cis}{\textit{cis}}

\newcommand{\re}{{\rm e}}
\renewcommand{\l}{{\rm lin}}

\newcommand{\schr}{Schr\"{o}dinger}
\newcommand{\Group}{\textit{\textbf{G}}}

\title{Theoretical rotation-vibration spectroscopy of {\it cis}- and {\it trans}-diphosphene (P$_2$H$_2$) and the deuterated species P$_2$HD}

\author{Alec Owens}
\email{alec.owens.13@ucl.ac.uk}
\affiliation{Department of Physics and Astronomy, University College London, Gower Street, WC1E 6BT London, United Kingdom}

\author{Sergei N. Yurchenko}
\email{s.yurchenko@ucl.ac.uk}
\affiliation{Department of Physics and Astronomy, University College London, Gower Street, WC1E 6BT London, United Kingdom}

\date{\today}

\begin{abstract}
Growing astronomical interest in phosphorous (P) chemistry is stimulating the search for new interstellar P-bearing molecules; a task requiring detailed knowledge of the microwave and infrared molecular spectrum. In this work, we present comprehensive rotation-vibration line lists of the \cis- and \trans-isomers of diphosphene (P$_2$H$_2$). The line lists have been generated using robust, first-principles methodologies based on newly computed, high-level \ai\ potential energy and dipole moment surfaces. Transitions are considered between states with energies up to $8000$~cm$^{-1}$ and total angular momentum $J\leq25$. These are the first-ever line lists to be reported for P$_2$H$_2$ and they should significantly facilitate future spectroscopic characterization of this system. The deuterated species \trans-P$_2$HD and the effect of its dynamic dipole moment on the rovibrational spectrum is also discussed.
\end{abstract}

\pacs{}

\maketitle 

\section{Introduction}
\label{sec:intro}

Phosphorous (P) chemistry is of increasing astronomical interest~\citep{JimnezSerra2018}, notably because of the essential role phosphorous plays in biochemical processes. To date, only a handful of P-bearing molecules (CP~\citep{CP:Guelin1990}, PN~\citep{PN:Turner1987,PN:Ziurys1987}, PO~\citep{PO:Tenenbaum2007}, C$_2$P~\citep{CCP:Halfen2008}, PH$_3$~\citep{PH3:Agundez2014}, HCP~\citep{HCP:Agundez2007}, and the tentative identification of NCCP~\citep{NCCP:Agundez2014}) have been detected in interstellar and circumstellar environments, and such observations require detailed knowledge of the microwave and infrared molecular spectrum. Despite its relatively low cosmic abundance, the fact that phosphorous is highly reactive means that other P-bearing species are likely to be observed in the future.

One potential system is diphosphene (P$_2$H$_2$), which structurally can exist in either the \cis- or \trans-isomeric form. Diphosphene has been detected in the pyrolysis of diphosphine (P$_2$H$_4$)~\cite{Fehlner1966}, and is also a potential intermediate in the photolysis of phosphine (PH$_3$) to phosphorus and hydrogen~\cite{Ferris1980}. The phosphorus-phosphorus double bond in P$_2$H$_2$, which is an unusual property in phosphorous chemistry, has led to interest in this molecule and a number of theoretical studies have been reported~\cite{Yoshifuji1983.P2H2,Ha1984.P2H2,Allen1985.P2H2,
Schmidt1986.P2H2,
Ito1986.P2H2,Allen1986.P2H2,Nguyen1989.P2H2,
Fueno1995.P2H2,Matus2007.P2H2,
09LuSiEv.P2H2,10LuHaSi.P2H2,VogtGeisse2012.P2H2}. However, they have largely been concerned with the calculation of isomerization transition states and the respective equilibrium structures. As such, spectroscopic data on this system is in very short supply. For example, we are unaware of any known values for the fundamental vibrational frequencies of the \hpph\ isomers. We thus find it worthwhile to investigate the rovibrational spectrum of \cis- and \trans-P$_2$H$_2$ using rigorous, first-principles methodologies.

\begin{figure}
\centering
\includegraphics{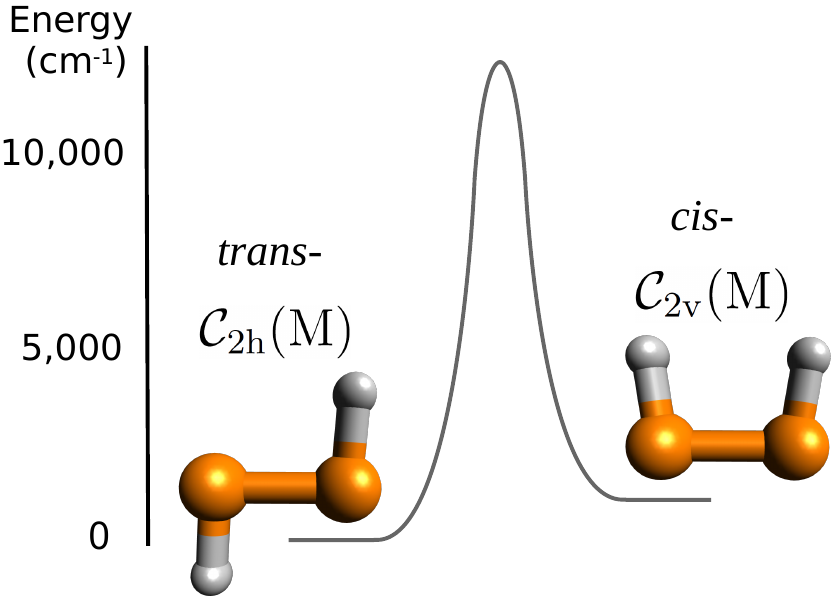}
\caption{\label{fig:cis_trans}Illustration of the \cis- and \trans-isomers of \hpph\ and the potential energy barrier between the two structures. In our approach, the isomers are treated as separate molecules with different molecular symmetry. Note that other barriers and isomers, namely diphosphinylidene (PPH$_2$), are not shown.}
\end{figure}

Nowadays, the rovibrational spectra of small, closed-shell polyatomic molecules can be accurately calculated using high-level \ai\ methods in conjunction with a variational treatment of nuclear motion~\citep{Tennyson:2016}. Purely \ai\ potential energy surfaces (PESs) can predict vibrational energy levels to within $\pm$1~cm$^{-1}$~\citep{Polyansky:Science299:539,Schwenke:SpecActA58:849,
YaYuRi11.H2CS,Malyszek:JCC34:337,15OwYuYa.CH3Cl,15OwYuYa.SiH4,16OwYuYa.CH4,19OwYaKu.CH3F}, while \ai\ dipole moment surfaces (DMSs) are capable of producing transition intensities comparable to, if not more reliable in some instances, than experiment~\cite{13Yuxxxx.method,Tennyson:JMolSpec298:1}. Such approaches have considerable predictive power and are becoming commonplace in a range of applications, for example, the ExoMol database~\citep{ExoMol2012,ExoMol2016} is constructing molecular line lists to aid the atmospheric characterization of exoplanets and other hot bodies. It is within the ExoMol computational framework~\citep{ExoSoft2016} that we treat diphosphene. As shown in Fig.~\ref{fig:cis_trans}, the potential energy barrier between \cis- and \trans-\hpph\ is large enough (approximately $h c \cdot 12\,300$~cm$^{-1}$~\citep{09LuSiEv.P2H2}, where $h$ is the Planck constant and $c$ is the speed of light) that we can consider them as separate molecules with different molecular symmetry.

The paper is organized as follows: The electronic structure calculations and analytic representation of the PESs and DMSs are described in Sec.~\ref{sec:PES} and Sec.~\ref{sec:DMS}, respectively. The variational nuclear motion computations and intensity simulations are detailed in Sec.~\ref{sec:trove}. The line lists, which consider transitions up to $J=25$ in the 0--8000~cm$^{-1}$ region, are presented in Sec.~\ref{sec:results} as well as a discussion of the deuterated species P$_2$HD and the effect of its dynamic dipole moment on the rovibrational spectrum. Concluding remarks are offered in Sec.~\ref{sec:conc}.

\section{Potential Energy Surface}
\label{sec:PES}

\subsection{Electronic structure calculations}

  The {\cis} and {\trans}-isomers of \hpph\ are known to have little multireference character~\citep{09LuSiEv.P2H2}, so their electronic structure can be accurately described using coupled cluster methods. Similar to our previous work on \textit{ab initio} PESs~\citep{15OwYuYa.CH3Cl,15OwYuYa.SiH4,16OwYuYa.CH4,19OwYaKu.CH3F}, we employ focal-point analysis~\citep{Csaszar98} to represent the total electronic energy as
\begin{equation}\label{eq:tot_en}
E_{\mathrm{tot}} = E_{\mathrm{CBS}}+\Delta E_{\mathrm{CV}}+\Delta E_{\mathrm{SR}}+\Delta E_{\mathrm{HO}} .
\end{equation}
The energy at the complete basis set (CBS) limit $E_{\mathrm{CBS}}$ was determined using the explicitly correlated coupled cluster method CCSD(T)-F12b~(Ref.~\onlinecite{Adler07}) with the F12-optimized correlation consistent basis sets, cc-pVTZ-F12 and cc-pVQZ-F12~\cite{Peterson08}. A parameterized, two-point formula~\cite{Hill09}
\begin{equation}\label{eq:cbs_extrap}
E^{C}_{\mathrm{CBS}} = (E_{n+1} - E_{n})F^{C}_{n+1} + E_{n} ,
\end{equation}
was used to extrapolate to the CBS limit. The coefficient $F^{C}_{n+1}$ is unique to the CCSD-F12b and (T) component of the total correlation energy and assumed a value of $F^{\mathrm{CCSD-F12b}}=1.363388$ and $F^{\mathrm{(T)}}=1.769474$, respectively~\citep{Hill09}. No extrapolation was applied to the Hartree-Fock (HF) energy, instead the HF+CABS (complementary auxiliary basis set) singles correction~\cite{Adler07} computed in the quadruple zeta basis set was used. Calculations employed the frozen core approximation and the diagonal fixed amplitude ansatz 3C(FIX)~\cite{TenNo04} with a Slater geminal exponent value of $\beta=1.0$~$a_0^{-1}$~\cite{Hill09}. For the auxiliary basis sets (ABS), the resolution of the identity OptRI~\cite{Yousaf08} basis was employed, and the cc-pV5Z/JKFIT~\cite{Weigend02} and aug-cc-pwCV5Z/MP2FIT~\cite{Hattig05} basis sets for density fitting. Calculations were performed with MOLPRO2015~\cite{Werner2012} unless stated otherwise.

 Core-valence (CV) electron correlation $\Delta E_{\mathrm{CV}}$ was accounted for at the CCSD(T)-F12b/cc-pCVTZ-F12~(Ref.~\onlinecite{Hill10}) level of theory with the same ansatz and ABS as listed above but with a Slater geminal exponent value of $\beta=1.4$~$a_0^{-1}$. The (1\textit{s}) orbital of the phosphorous atoms was frozen in all-electron calculations since basis sets cannot describe this orbital adequately. 
 
 Scalar relativistic (SR) effects $\Delta E_{\mathrm{SR}}$ were computed using the second-order Douglas-Kroll-Hess approach~\cite{dk1,dk2} at the CCSD(T)/cc-pVQZ-DK~(Ref.~\onlinecite{dk_basis}) level of theory in the frozen core approximation. After the completion of this study it was brought to our attention that the electronic energies that include the Douglas-Kroll one-electron integrals lose accuracy when using a basis set above triple-zeta quality, see e.g. Ref.~\onlinecite{Huang2008}. Tests were performed on some select geometries to investigate this degradation but the results were inconclusive. Given that the size of the SR correction ranged from $\approx\pm 350$~cm$^{-1}$, we do not expect this effect to significantly change the shape of the PESs and the subsequent accuracy of our variational calculations.
 
 Higher-order (HO) correlation effects were computed using the hierarchy of coupled cluster methods CCSD(T), CCSDT, and CCSDT(Q), such that $\Delta E_{\mathrm{HO}} = \Delta E_{\mathrm{T}} + \Delta E_{\mathrm{(Q)}}$. Here, the full triples contribution is $\Delta E_{\mathrm{T}} = E_{\mathrm{CCSDT}}-E_{\mathrm{CCSD(T)}}$, and the perturbative quadruples contribution is $\Delta E_{\mathrm{(Q)}} = E_{\mathrm{CCSDT(Q)}}-E_{\mathrm{CCSDT}}$. Calculations in the frozen core approximation employed the general coupled cluster approach~\cite{Kallay05,Kallay08} as implemented in the MRCC code~\cite{mrcc} interfaced to CFOUR~\cite{cfour}. The correlation consistent basis sets cc-pVTZ(+d for P) and cc-pVDZ(+d for P)~\cite{Dunning89} were utilized for the full triples and perturbative quadruples contributions, respectively.

 The terms in Eq.~\eqref{eq:tot_en} were determined on a uniformly-spaced grid of 68,686 nuclear geometries with energies up to $h c \cdot 19\,000$~cm$^{-1}$. The grid was constructed in terms of six internal coordinates: the P{--}P bond length $1.6\leq r_{\rm PP}\leq 3.5$~$\mathrm{\AA}$, two P{--}H bond lengths $1.0\leq r_{\rm PH_{1}}, r_{\rm PH_{2}}\leq 3.0$~$\mathrm{\AA}$, two $\angle(\mathrm{PP}\mathrm{H}_i)$ interbond angles $40\leq \theta_{\rm PPH_{1}}, \theta_{\rm PPH_{2}}\leq 170^{\circ}$, and the dihedral angle $0 \leq \tau_{\rm HPPH}\leq 180^{\circ}$ between the planes containing the P{--}P{--}H$_1$ and P{--}P{--}H$_2$ nuclei. Geometries with $0 \leq \tau_{\rm HPPH}\leq 90^{\circ}$ were classified as  \cis-P$_2$H$_2$, while those with $90 < \tau_{\rm HPPH}\leq 180^{\circ}$ were treated as \trans-P$_2$H$_2$. This separation is consistent with the value of the dihedral angle corresponding to the isomerization transition state linking the \cis- and \trans-isomers~\citep{09LuSiEv.P2H2}. All \ai\ points computed for the PESs are provided as supplementary material.
 
  We mention that the energy difference between the equilibrium structures of \cis- and \trans-\hpph\ (considering CBS extrapolation and all HL corrections) was computed to be 1229.7~\cm, slightly higher than the \trans-\cis\ energy gap of $1119$~\cm\ determined in the theoretical study of Ref.~\onlinecite{09LuSiEv.P2H2}.

\subsection{Analytic representation}

The potential energy function of the \cis- and \trans-isomers was represented using the following expansion,
\begin{equation}
V =  \sum_{i_1,i_2,\ldots,i_6} f_{i_1,i_2,\ldots,i_6} \xi_1^{i_1} \xi_2^{i_2} \xi_3^{i_3} \xi_4^{i_4}\xi_5^{i_5}\xi_6^{i_6} ,
\end{equation}
with maximum expansion order $i_1+i_2+i_3+i_4+i_5+i_6=6$. This analytic representation is essentially a sextic force field in an appropriate coordinate system with proper limiting behaviour; an important factor in variational calculations. The coordinates,
\begin{eqnarray}
\label{eq:coords_pes}
  \xi_1 &=& 1 - e^{-(\alpha \Delta R)}, \\
  \xi_2 &=& 1 - e^{-(\beta \Delta r_1)}, \\
  \xi_3 &=& 1 - e^{-(\beta \Delta r_2)}, \\
  \xi_4 &=& \Delta \theta_1, \\
  \xi_5 &=& \Delta \theta_2, \\
  \xi_6 &=& \cos\tau-\cos\tau^{\rm ref},
\end{eqnarray}
where $R = r_{\rm PP}$, $r_1  = r_{\rm PH_{1}} $, $r_2  = r_{\rm PH_{2}} $, $\theta_1 = \theta_{\rm PPH_{1}}$, $\theta_2 = \theta_{\rm PPH_{2}}$, $\tau = \tau_{\rm HPPH}$, and the deviations $\Delta$ are taken relative to the respective reference equilibrium values $R^{\rm ref}$, $r_{1}^{\rm ref}=r_{2}^{\rm ref}$, $\theta_1^{\rm ref}=\theta_2^{\rm ref}$, and $\tau^{\rm ref}=0$ or 180$^{\circ}$ for the \cis\ and \trans\ configurations, respectively. The stretching Morse oscillator parameters $\alpha$ and $\beta$ were optimized in the fittings and had slightly different values for the \cis\ and \trans\ PESs, see the supplementary material.

The expansion parameters $f_{i_1,i_2,\ldots,i_6}$ were established through least-squares fittings to the \textit{ab initio} data and were subject to the permutation condition,
\begin{equation}
V(R,r_2,r_1,\theta_2,\theta_1,\tau) = V(R,r_1,r_2,\theta_1,\theta_2,\tau),
\end{equation}
i.e., $f_{i_1,i_3,i_2,i_5,i_4,i_6} = f_{i_1,i_2,i_3,i_4,i_5,i_6}$ for $i_2 \ne i_3$ or $i_4 \ne i_5$. The \trans-PES required 737 parameters which were determined from fitting to 36,609 points, while the \cis-PES needed 733 parameters fitted to 32,722 points. Weight factors of the form suggested by~ \citet{Schwenke97}
\begin{equation}\label{eq:weights}
w_i=\left(\frac{\tanh\left[-0.0006\times(\tilde{E}_i - 15{\,}000)\right]+1.002002002}{2.002002002}\right)\times\frac{1}{N\tilde{E}_i^{(w)}} ,
\end{equation}
were employed in the fit to favour energies below $15{\,}000{\,}$cm$^{-1}$. Here, $\tilde{E}_i^{(w)}=\max(\tilde{E}_i, 10{\,}000)$ where $\tilde{E}_i$ is the potential energy at the $i$th geometry above equilibrium and the normalization constant $N=0.0001$ (all values in cm$^{-1}$). Watson's robust fitting scheme~\cite{Watson03} was also utilized to reduce the weights of outliers and further improve the description at lower energies. 

\begin{figure}
\centering
\includegraphics{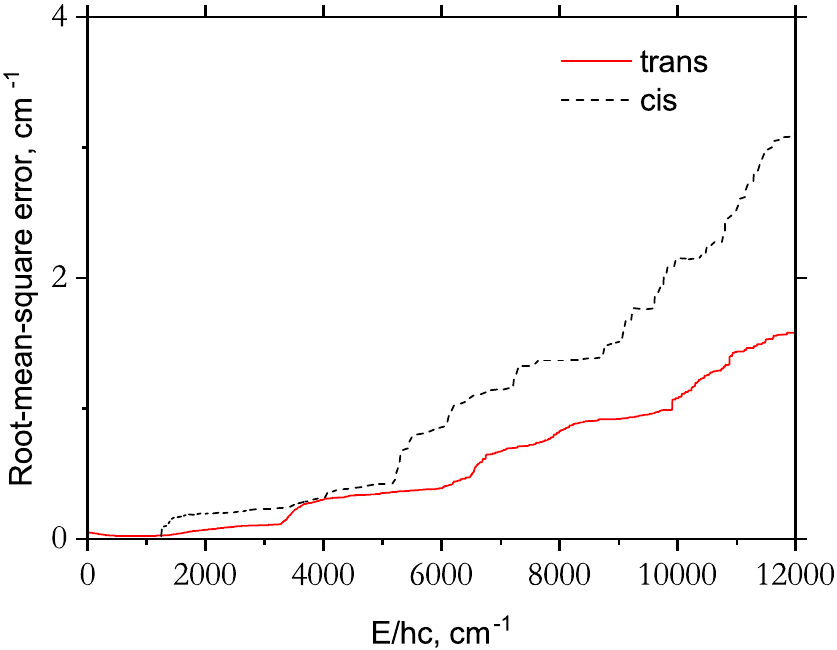}
\caption{\label{fig:rms}Root-mean-square errors of the PES fittings for \cis- and \trans-\hpph.}
\end{figure}

The results of the PES fittings are shown in Fig.~\ref{fig:rms} where it can be seen that the errors for the \cis-PES are systematically larger than the \trans-PES. Because the \cis-isomer resides at higher energy, this would explain the increased fitting error compared to \trans-\hpph. In both cases, the quality of the fit is excellent around the equilibrium region but gradually deteriorates as we go to higher energies. The isomerization transition state between \cis- and \trans-\hpph\ lies at around $h c \cdot 12\,300$~cm$^{-1}$ and requires a multireference description~\citep{09LuSiEv.P2H2}. As such, the computed coupled cluster energies will be unreliable as we sample the PES in the region of the isomerization transition state, and this will contribute to the degradation in the fittings, particularly for the \cis-isomer. However, this should not substantially affect our rovibrational computations as we are treating the \cis- and \trans-isomers as separate molecules with different PESs. The expansion parameters of the PESs are provided as supplementary material along with a program to construct the analytic representation.

\section{Dipole Moment Surface}
\label{sec:DMS}

\subsection{Electronic structure calculations}

The DMSs were generated on the same grid of nuclear geometries as the {\it cis-} and {\it trans-} PESs. Defining a Cartesian laboratory-fixed $XYZ$ coordinate system with origin at one of the P nuclei, an external electric field with components $\pm0.005$~a.u. was applied along each coordinate axis and the respective dipole moment component $\mu_A$ for $A=X,Y,Z$ determined through central finite differences. Calculations were performed at the CCSD(T)/aug-cc-pVTZ(+d for P) level of theory in the frozen core approximation using MOLPRO2015~\cite{Werner2012}.

\subsection{Analytic representation}

To represent the DMSs, it is beneficial to transform to a suitable molecule-fixed $xyz$ axis system before fitting an analytic representation to the \ai\ data. For P$_2$H$_2$, the following was used: The $z$ axis is aligned along the P{--}P bond, while the $x$ axis is perpendicular and lies in the plane bisecting the
two P{--}P{--}H planes (i.e., the planes containing the P{--}P and P{--}H bonds). The $y$ axis is oriented such that the $xyz$ axis system is right-handed. This representation is similar to that used for hydrogen peroxide (HOOH)~\cite{15AlOvPo.H2O2}.

For \trans-\hpph, the DMS spans the \Ch{2}(M) symmetry group with the $x$, $y$ and $z$ components of the dipole moment transforming according to the $A_{u}$, $B_{u}$, and $B_{u}$ irreducible representations, respectively. For \cis-\hpph, the DMS spans the \Cv{2}(M) symmetry group and the $x$, $y$ and $z$ components transform according to the $A_{1}$, $B_{1}$, and $B_{2}$ representations, respectively. In the molecule-fixed axis system, the symmetry-adapted projections of the electronically averaged Cartesian dipole moment components $\mu_{x}, \mu_{y}$ and $\mu_{z}$ are given as analytic representations, where each component is expanded in Taylor series around the equilibrium configuration in terms of internal coordinates, that is,
\begin{eqnarray}
  \bar\mu_{x} &=& \cos(\tau/2) \sum_{i_1,i_2,\ldots,i_6} F_{i_1,i_2,\ldots,i_6}^{(x)} \zeta_1^{i_1} \zeta_2^{i_2} \zeta_3^{i_3} \zeta_4^{i_4}\zeta_5^{i_5}\zeta_6^{i_6}, \\
  \bar\mu_{y} &=& \sin(\tau/2) \sum_{i_1,i_2,\ldots,i_6} F_{i_1,i_2,\ldots,i_6}^{(y)} \zeta_1^{i_1} \zeta_2^{i_2} \zeta_3^{i_3} \zeta_4^{i_4}\zeta_5^{i_5}\zeta_6^{i_6}, \\
  \bar\mu_{z} &=&  \sum_{i_1,i_2,\ldots,i_6} F_{i_1,i_2,\ldots,i_6}^{(z)} \zeta_1^{i_1} \zeta_2^{i_2} \zeta_3^{i_3} \zeta_4^{i_4}\zeta_5^{i_5}\zeta_6^{i_6}.
\end{eqnarray}
The coordinates are defined as,
\begin{eqnarray}
  \zeta_1 &=& \Delta R  \, e^{-(\Delta R)^2}, \\
  \zeta_2 &=& \Delta r_{1} \,  e^{-(\Delta r_1)^2}, \\
  \zeta_3 &=& \Delta r_{2} \,  e^{-(\Delta r_2)^2}, \\
  \zeta_4 &=& \cos\theta_1-\cos\theta_1^{\rm ref}, \\
  \zeta_5 &=& \cos\theta_2-\cos\theta_2^{\rm ref}, \\
  \zeta_6 &=& \cos\tau-\cos\tau^{\rm ref},
\end{eqnarray}
and the expansion parameters of the $x$, $y$ and $z$ dipole moment components obey the following permutation rules,
\begin{eqnarray}
\label{e:mux}
  F_{i_1,i_2,i_3,i_4,i_5,i_6}^{(x)} &=&  F_{i_1,i_3,i_2,i_5,i_4,i_6}^{(x)},\\
  \label{e:muy}
 F_{i_1,i_2,i_3,i_4,i_5,i_6}^{(y)} &=& -F_{i_1,i_3,i_2,i_5,i_4,i_6}^{(y)},\\
  \label{e:muz}
 F_{i_1,i_2,i_3,i_4,i_5,i_6}^{(z)} &=& -F_{i_1,i_3,i_2,i_5,i_4,i_6}^{(z)},
\end{eqnarray}
which corresponds to permuting the two hydrogen atoms. Thus, $F_{0,0,0,0,0,i_6}^{(\alpha)} = 0$ when $\alpha=y,z$ for any value of the $i_6$ index.

\begin{figure*}
\centering
\includegraphics{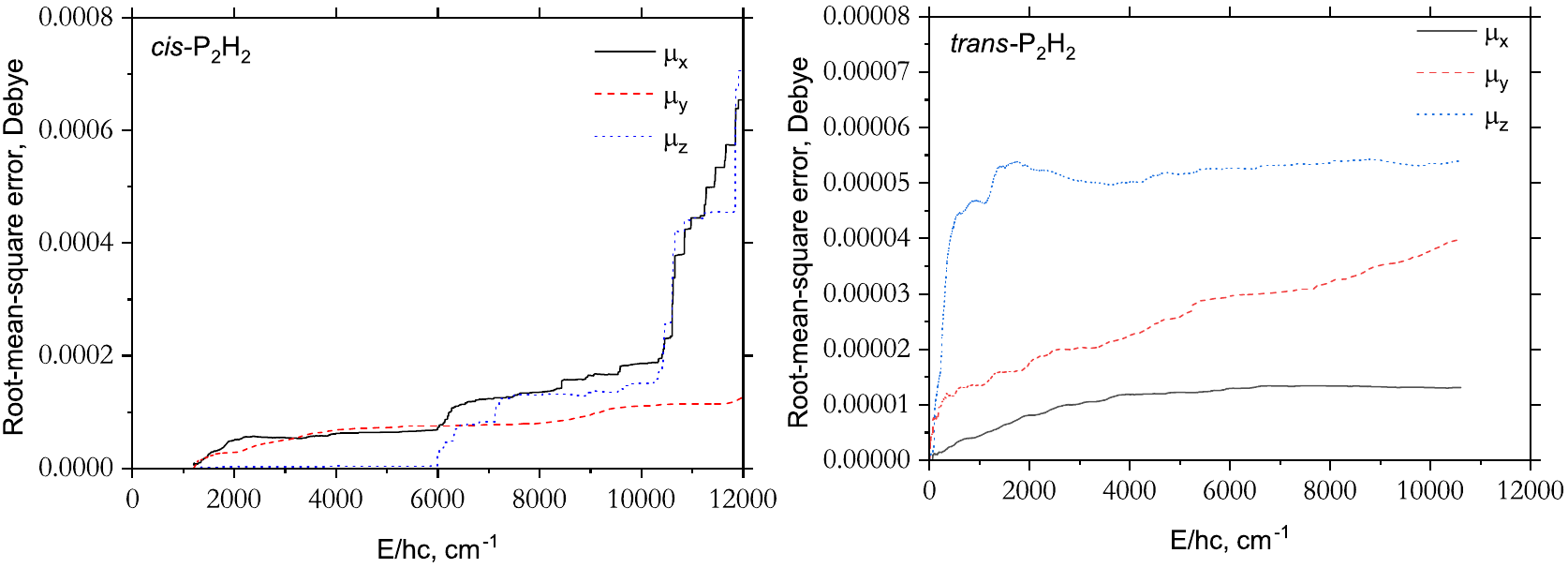}
\caption{\label{fig:rms:dms}Root-mean-square errors of the DMS fittings for \cis- and \trans-\hpph.}
\end{figure*}

 The expansion coefficients $F^{(\alpha)}_{i_1,i_2,\ldots}$ for $\alpha=x,y,z$ were determined simultaneously through a least-squares fitting to the \textit{ab initio} data. Watson's robust fitting scheme~\citep{Watson03} and the same weight factors as shown in Eq.~\eqref{eq:weights} were used. Similarly, the reference equilibrium parameters assumed similar values to those employed in the PES fittings, see the supplementary material. For the \cis-isomer, the three dipole moment surfaces, $\mu_x$, $\mu_y$, and $\mu_z$, required 378, 242 and 404 parameters, respectively, whilst the \trans\ dipole components used 496, 420 and 421 parameters. The results of the DMS fittings are illustrated in Fig.~\ref{fig:rms:dms}. The $\mu_x$ component of the \cis-isomer (which has a permanent electric dipole) has the largest magnitude and hence the largest  errors, otherwise the behaviour of the DMS fittings is as expected. The expansion parameters of the DMSs are given as supplementary material along with a program to construct the analytic representation.

\section{Variational calculations}
\label{sec:trove}

 The computer program TROVE~\citep{TROVE2007} was used for all nuclear motion computations. Since its procedures and methodology are well documented~\cite{TROVE2007,YuBaYa09.NH3,15YaYu.ADF,Symmetry:2017}, we summarize only the key aspects relevant for this work. Calculations are based on the assumption that the two isomers are separated by an impenetrable barrier ($\approx h c \cdot 12\,300$~cm$^{-1}$) with zero tunneling. The \cis- and \trans-isomers are treated as separate molecules with different molecular symmetry.

 The rovibrational Hamiltonian was constructed numerically with both the kinetic and potential energy operators truncated at 6th order (see Refs.~\onlinecite{TROVE2007,15YaYu.ADF} for a discussion of the associated errors of such a scheme). The Hamiltonian was represented in the Eckart frame as a power series expansion around the equilibrium geometry in terms of the six coordinates introduced in Eqs.~\eqref{eq:coords_pes}--(9), except in the case of the kinetic energy operator which used linear displacement variables $(r_i - r_i^{\mathrm{ref}})$ for the stretching terms. The symmetrized rovibrational basis set was constructed using a general, multi-step contraction scheme~\cite{Symmetry:2017}. The polyad number, defined as 
\begin{equation}
\label{eq:polyad}
P = n_1 + n_2 + n_3 + n_4 + n_5 + n_6 ,
\end{equation}
was used to control the size of the primitive and contracted basis sets. Here, the quantum numbers $n_k$ for $k=1,\ldots,6$ correspond to primitive basis functions $\phi_{n_k}$, which are determined by solving one-dimensional Schr\"{o}dinger equations for each $k$th vibrational mode. Products of $\phi_{n_k}$ are used to build the vibrational basis set. Initially, reduced symmetry sub-spaces are set up by coupling equivalent modes and solving the reduced-mode Schr\"odinger equations; for example all stretching degrees of freedom are treated together. The resulting eigenfunctions are then combined and used as a basis set to solve the full-dimensional problem. The final $J=0$ basis sets were truncated with $P \le 12$ and $P \le 14$, resulting in 8892 and 8701 basis functions (containing all symmetries) for \cis- and \trans-\hpph, respectively. All primitive basis functions were generated using the Numerov-Cooley approach~\cite{Numerov1924,Cooley1961} on grids of 1000 points and then symmetrized using the TROVE symmetry-adaptation procedure \cite{Symmetry:2017}. The $J=0$ eigenfunctions were multiplied with symmetrized rigid-rotor functions to produce the final basis set for $J>0$ calculations. Nuclear masses were employed in TROVE calculations.

 The \trans-isomer is of \Ch{2}(M) symmetry with transitions obeying the dipole selection rules $A_g \leftrightarrow A_u, B_g \leftrightarrow B_u$. The nuclear spin statistical weights are $g_{\mathrm{ns}}=\lbrace 10,10,6,6\rbrace$ for states of symmetry $\lbrace A_g,A_u,B_g,B_u\rbrace$, respectively. In calculations, the Hamiltonian was expanded around equilibrium values of $r_{\rm PP}=2.025$~$\mathrm{\AA}$, $r_{\rm PH}=1.419$~$\mathrm{\AA}$, $\theta_{\rm PPH}=93.88^{\circ}$ and $\tau_{\rm HPPH}=180.0^{\circ}$. The zero-point energy (ZPE) was computed to be 3865.759~\cm.

 The \cis-isomer is of \Cv{2}(M) symmetry with transitions obeying the dipole selection rules $A_1 \leftrightarrow A_2, B_1 \leftrightarrow B_2$. The nuclear spin statistical weights are $g_{\mathrm{ns}}=\lbrace 10,10,6,6\rbrace$ for states of symmetry $\lbrace A_1,A_2,B_1,B_2\rbrace$, respectively. In calculations, the Hamiltonian was expanded around equilibrium values of $r_{\rm PP}=2.041$~$\mathrm{\AA}$, $r_{\rm PH}=1.419$~$\mathrm{\AA}$, $\theta_{\rm PPH}=98.89^{\circ}$ and $\tau_{\rm HPPH}=0.0^{\circ}$. The ZPE was computed to be 5025.816~\cm, i.e., 1160.061~\cm\ above that of the \trans-isomer.

 All transitions and Einstein $A$ coefficients involving states with rotational excitation up to $J=25$ have been computed using the MPI-GAIN program~\cite{GAIN-MPI:2017} for the 0--8000~\cm\ region (above the ZPE of each isomer). The lower and upper state energy thresholds were chosen to be 4000~\cm\ and 8000~cm$^{-1}$, respectively. Spectral simulations employed the ExoCross code~\cite{Exocross:2018} and were carried out at a temperature of $T=296$~K with a partition function value of $Q=50\,984.9$ and $Q=50\,503.2$ for \cis- and \trans-\hpph, respectively. Transitions obeyed the symmetry selection rules defined above and the standard rotational selection rules $J\p-J\pp=0,\pm 1,\; J\p+J\pp \ne 0$, where $\p$ and $\pp$ denote the upper and lower state, respectively. Overall, the \trans-\hpph\ line list contained 10,667,208,951 transitions between 5,881,876 states, while the \cis-\hpph\ line list consisted of 11,020,092,365 transitions between 6,009,302 states. 

As an aside, we have found it worthwhile to investigate the effects of isotopic substitution on the spectra of diphosphene, in particular the rotational spectrum of the \trans-isomer. To this end, variational calculations for the deuterated species \trans-P$_2$HD have been performed utilizing the \trans-\hpph\ PES and DMS. Since this is an asymmetric species, calculations employed the ${\mathcal C}_{\rm s}$(M) molecular symmetry group which consists of only two irreducible representations, $A^{\prime}$ and $A^{\prime\prime}$, therefore leading to larger symmetry-adapted Hamiltonian matrices when solving the respective Schr\"{o}dinger equations with TROVE. The contracted vibrational basis set used a polyad truncation number of $P \le 12$ resulting in 17\,312 functions, significantly larger than those of the P$_2$H$_2$ isomers. Due to this increased size, our study of P$_2$HD was restricted to states with $J\le 10$. The final line list contained 2,499,185,732 transitions between 895,884 states in the 0--8000~\cm\ region. Note that the nuclear spin statistical weight $g_{\rm ns}=24$ for P$_2$HD.

\section{Results}
\label{sec:results}

\subsection{Vibrational $J=0$ energies}

The computed fundamental frequencies of \cis- and \trans-\hpph\ are listed in Table~\ref{tab:j0_cis} and \ref{tab:j0_trans}, respectively. We are unaware of any known values to compare against, however, our computed ``anharmonic'' values are in sensible agreement with accurate \ai\ harmonic frequencies computed using coupled cluster and multireference methods in conjunction with large Gaussian basis sets~\citep{09LuSiEv.P2H2}. Based on our previous experience of generating high-level \ai\ PESs of closed-shell molecules~\citep{15OwYuYa.CH3Cl,15OwYuYa.SiH4,16OwYuYa.CH4,19OwYaKu.CH3F}, which often utilized the same focal-point approach and levels of theory, the combined rms error of the computed fundamentals when compared to experiment has been well within $\pm 1$~cm$^{-1}$. The fundamentals of P$_2$HD are listed in Table~\ref{tab:j0_p2hd}. 

Convergence tests utilizing larger vibrational basis sets ($P \le 16$ in Eq.~\eqref{eq:polyad}) were performed for \cis- and \trans-\hpph. The convergence error for $J=0$ states below 3000~cm$^{-1}$ (used in the line lists) showed a mean absolute deviation of 0.14~cm$^{-1}$ for \cis-\hpph, and 0.02~cm$^{-1}$ for \trans-\hpph, with the fundamentals exhibiting orders-of-magnitude better convergence than other states. The improved convergence for \trans-\hpph\ is due to the larger polyad number in variational calculations. Naturally, the convergence errors increase at higher energies, particularly for highly excited modes which are the most difficult to converge. In the worst instances this error can be of the order of tens of wavenumbers. Some caution should therefore be exercised when using the line lists for transitions above 4000~cm$^{-1}$ (as a conservative estimate). A list of $J=0$ energies up to 8000~cm$^{-1}$ (above the ZPE of each isomer) with symmetry and TROVE quantum number labelling (determined from the contribution of the primitive basis functions) is provided as supplementary material.
 
\begin{table}
\tabcolsep=5pt
\caption{\label{tab:j0_cis}Computed fundamental term values (in cm$^{-1}$) of {\it cis-}P$_2$H$_2$.}
\begin{center}
	\begin{tabular}{l c l r}
	\hline\hline
 	 Mode & Sym. & Description & Calculated\\
	\hline	
     $\nu_1$ & $A_1$ & Symmetric PH stretching & 2290.93 \\[-1mm]
     $\nu_2$ & $A_1$ & Symmetric bending &  733.68 \\[-1mm]
     $\nu_3$ & $A_1$ & PP stretching &  593.87 \\[-1mm]
     $\nu_4$ & $A_2$ & Torsional motion &  678.21 \\[-1mm]
     $\nu_5$ & $B_2$ & Asymmetric PH stretching & 2285.13 \\[-1mm]
     $\nu_6$ & $B_2$ & Asymmetric bending &  825.70 \\[-1mm]
    \hline\hline
    \end{tabular}
\end{center}
\end{table}

\begin{table}
\tabcolsep=5pt
\caption{\label{tab:j0_trans}Computed fundamental term values (in cm$^{-1}$) of {\it trans-}P$_2$H$_2$.}
\begin{center}
	\begin{tabular}{l c l r}
	\hline\hline
 	 Mode & Sym. & Description & Calculated \\
	\hline	
     $\nu_1$ & $A_g$ & Symmetric PH stretching & 2263.88 \\[-1mm]
     $\nu_2$ & $A_g$ & Symmetric bending &  954.73 \\[-1mm]
     $\nu_3$ & $A_g$ & PP stretching &  604.52 \\[-1mm]
     $\nu_4$ & $A_u$ & Torsional motion &  753.94 \\[-1mm]
     $\nu_5$ & $B_u$ & Asymmetric PH stretching & 2280.89 \\[-1mm]
     $\nu_6$ & $B_u$ & Asymmetric bending &  669.07 \\[-1mm]
    \hline\hline
    \end{tabular}
\end{center}
\end{table}

\begin{table}
\tabcolsep=5pt
\caption{\label{tab:j0_p2hd}Computed fundamental term values (in cm$^{-1}$) of {\it trans-}P$_2$HD.}
\begin{center}
	\begin{tabular}{l c l r}
	\hline\hline
 	 Mode & Sym. & Description & Calculated \\
	\hline	
     $\nu_1$ & $A^{\prime}$ & PH stretching & 2233.04 \\[-1mm]
     $\nu_2$ & $A^{\prime}$ & PD stretching & 1652.41 \\[-1mm]
     $\nu_3$ & $A^{\prime}$ & PPH bending   &  871.69 \\[-1mm]
     $\nu_4$ & $A^{\prime}$ & PP stretching &  607.28 \\[-1mm]
     $\nu_5$ & $A^{\prime}$ & PPD bending   &  530.90 \\[-1mm]
     $\nu_6$ & $A^{\prime\prime}$ & Torsional motion &  658.04 \\[-1mm]
    \hline\hline
    \end{tabular}
\end{center}
\end{table}

\subsection{Line lists}

\begin{figure*}
\centering
\includegraphics{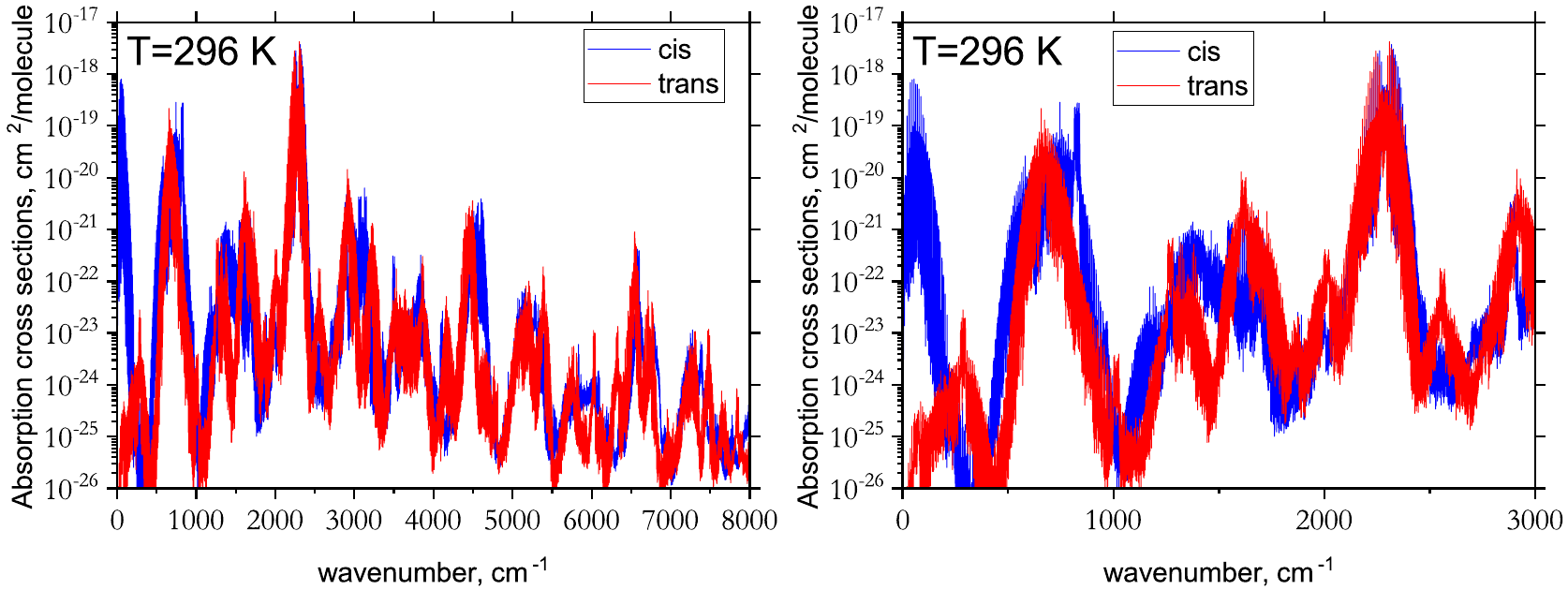}
\caption{\label{fig:overview}Overview of the rotation-vibration spectrum of \cis- and \trans-\hpph\ at $T=296$~K modelled using a Gaussian line profile with a half-width at half-maximum (HWHM) of 0.5~\cm. The right panel shows the 0--3000~\cm wavenumber region containing the fundamental bands. No scaling has been applied to either spectrum (see text).}
\end{figure*}

\begin{figure*}
\centering
\includegraphics{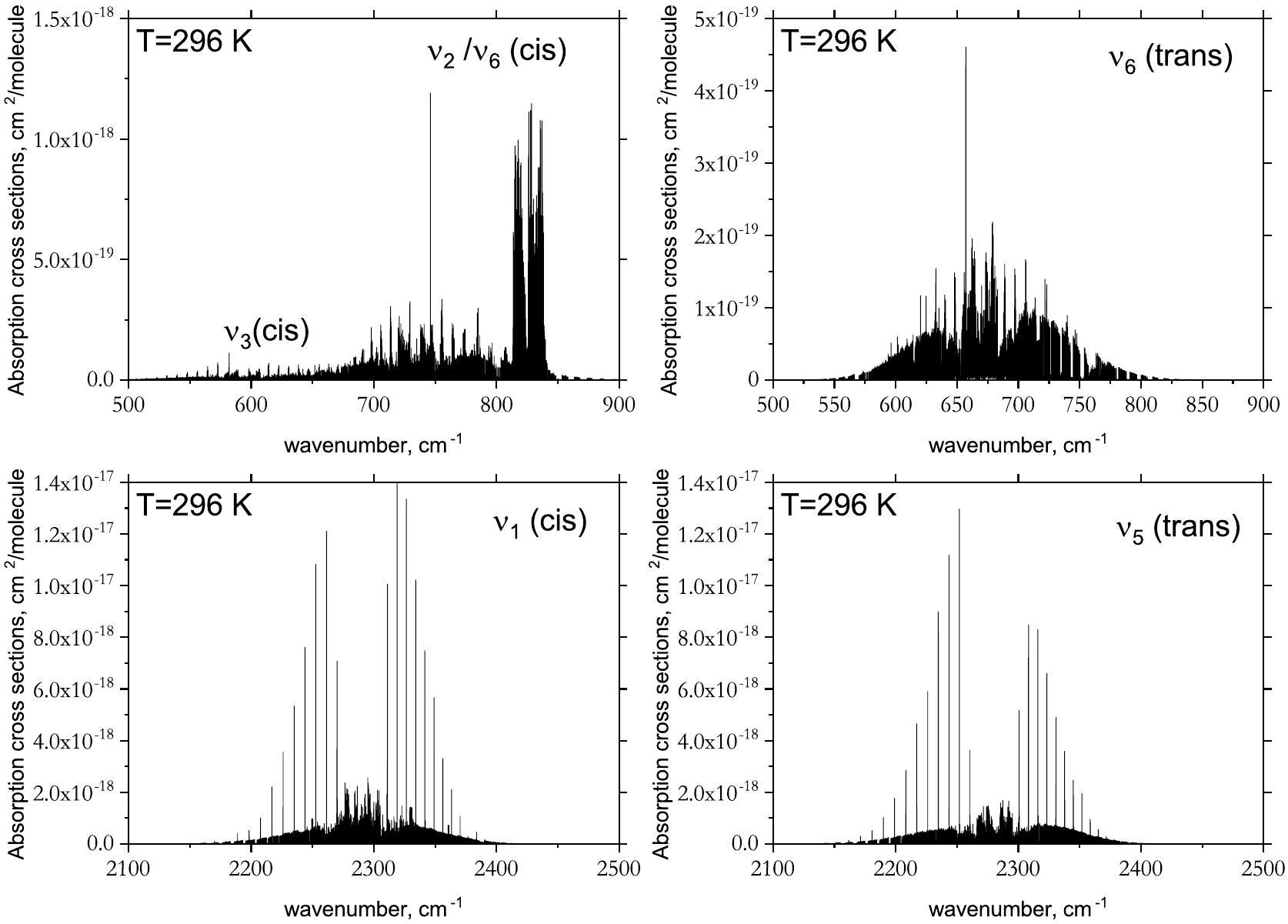}
\caption{\label{fig:nu1}Closer inspection of the fundamental bands of \cis- and \trans-\hpph\ at $T=296$~K modelled using a Gaussian line profile with a HWHM of 0.1~\cm.}
\end{figure*}

\begin{figure}
\centering
\includegraphics{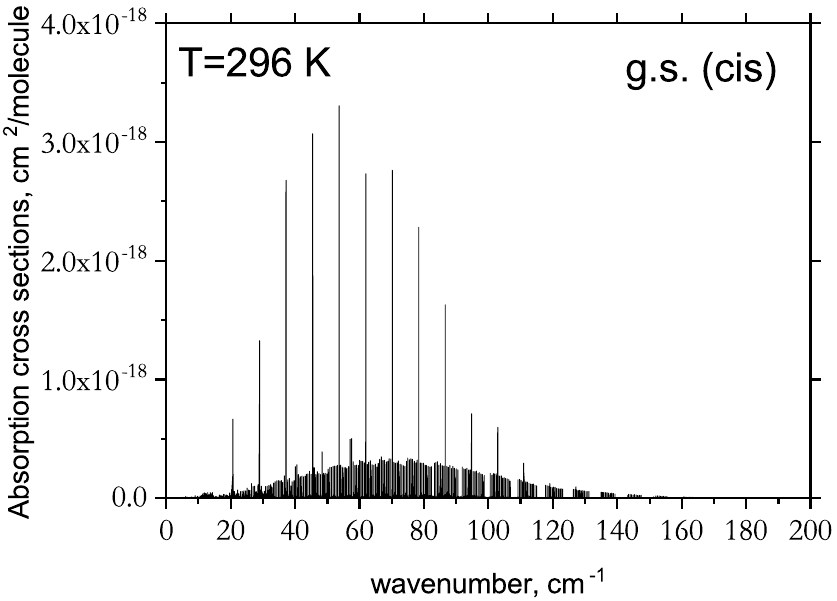}
\caption{\label{fig:gs}The rotational band of \cis-\hpph\ at $T=296$~K modelled using a Gaussian line profile with a HWHM of 0.1~\cm.}
\end{figure}

An overview of the rotation-vibration spectrum of \cis- and \trans-\hpph\ is shown in Fig.~\ref{fig:overview}. The strongest intensity features in both isomers occur for the PH stretching modes around $2300$~\cm. A closer inspection of these bands along with the other fundamentals is shown in Fig.~\ref{fig:nu1}. The most noticeable difference between the isomers occurs below 500~\cm. Because \trans-\hpph\ has no permanent dipole moment it possesses a very weak rotational spectrum. The \cis-isomer, however, has a large permanent dipole moment (computed in this study to be 1.033~D which is consistent with values determined in Ref.~\onlinecite{09LuSiEv.P2H2}) and a well pronounced rotational structure, evident in Fig.~\ref{fig:gs}.

\begin{figure}
\centering
\includegraphics{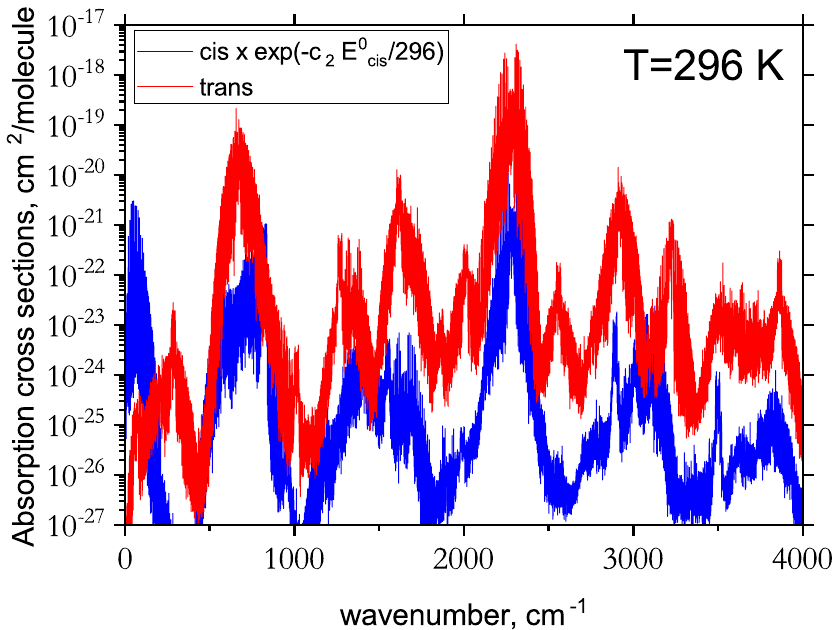}
\caption{\label{fig:Boltz}Spectra of \cis- and \trans-\hpph\ at $T=296$~K with the \cis-spectrum scaled using the Boltzmann factor $\exp(-hc\tilde{E}_{\rm ZPE}^{\rm \cis}/kT)$, where $\tilde{E}_{\rm ZPE}^{\rm \cis}=1160.061$~\cm.}
\end{figure}

The \cis- and \trans-isomers are relatively stable and the rate for isomerization is as low as $3\times10^{-12}$~s$^{-1}$ (corresponding to a half-life of $3\times10^{11}$~s) at $T=298$~K~\citep{09LuSiEv.P2H2}. Since the \cis-isomer has a higher ZPE, if both species were to exist in a region at local thermodynamic equilibrium, the spectrum of \cis-\hpph\ would appear slightly weaker in comparison to \trans-\hpph. In Fig.~\ref{fig:Boltz}, we show the relative temperature effects on the spectrum of \cis-\hpph\ at $T=296$~K using the Boltzmann scaling factor $\exp(-hc\tilde{E}_{\rm ZPE}^{\rm \cis}/kT)$ where $\tilde{E}_{\rm ZPE}^{\rm \cis}=1160.061$~\cm. Promisingly, the rotational band of \cis-\hpph\ appears strong enough to be detected. Note that no Boltzmann scaling was applied to the spectra in Fig.~\ref{fig:overview}.

\begin{figure}
\centering
\includegraphics{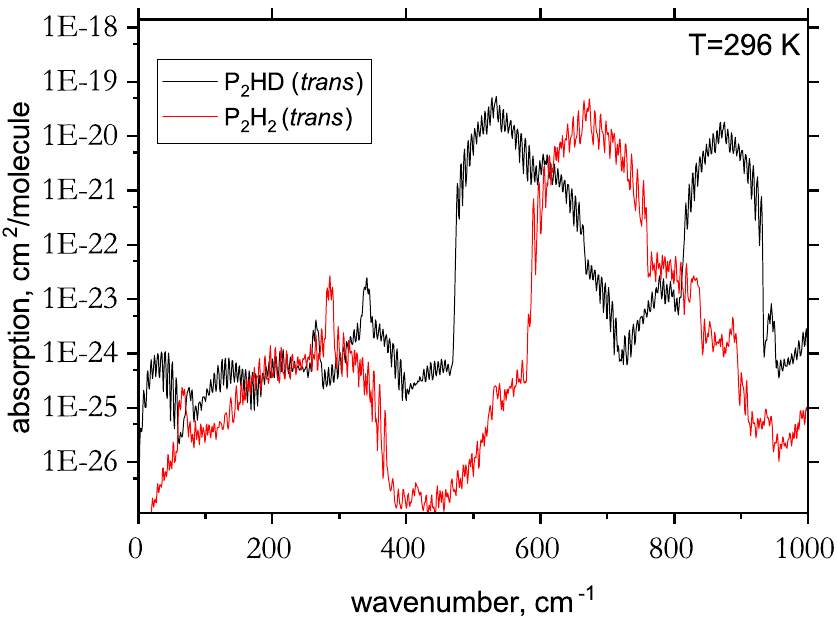}
\caption{\label{fig:HPPD}Spectrum of \trans-P$_2$HD compared to \trans-P$_2$H$_2$ at $T=296$~K modelled using a Gaussian line profile with a HWHM of 0.5~\cm.}
\end{figure}

In Fig.~\ref{fig:HPPD}, the spectrum of \trans-P$_2$HD is plotted alongside that of \trans-\hpph. The effect of isotopic substitution is clearly seen in the shifted band centers but more noticeably in the rotational spectrum below 200~cm$^{-1}$. Despite the fact that P$_2$HD has no permanent \ai\ dipole moment, the interaction of the rotational and  vibrational degrees of freedom in this asymmetric isomer leads to a non-zero effective dipole moment, thus increasing the intensity of the pure rotational lines by about three orders of magnitude. Although fairly weak the rotational spectrum of P$_2$HD should be detectable, at least in the laboratory. It is worth mentioning that recent work has investigated the effects of H~$\rightarrow$~D isotopic substitution on phosphine spectra~\cite{Viglaska2018}.

The computed line lists of these three species are available in the ExoMol format~\citep{ExoMol2016} from the ExoMol database at \url{www.exomol.com}. If combining the separate \cis- and \trans-\hpph\ line lists into one diphosphene spectrum, it is necessary to shift the ZPE of the \cis\ energy levels by 1160.061~\cm. 

Commenting on the accuracy of the line lists, we expect band centers of the fundamentals to be accurate to within $\pm$3~cm$^{-1}$ (as a conservative estimate) but this could be in the $\pm$1~cm$^{-1}$ range for certain bands. Line intensities should on average be accurate to within 5--10\% of experiment. These estimates are based on our previous experience generating high-level \ai\ PESs and DMSs for closed-shell systems using similar levels of theory~\citep{15OwYuYa.CH3Cl,15OwYuYa.SiH4,16OwYuYa.CH4,19OwYaKu.CH3F,16OwYuYa.CH3Cl}. That said, the isomerization transition state between \cis- and \trans-\hpph\ is located at around $h c \cdot 12\,300$~cm$^{-1}$ so the \ai\ points computed at geometries near this will be unreliable due to its multireference character. Also, because we have neglected the double-well nature of the PES of \hpph, computed higher rovibrational energies, i.e. those above $\approx 6000$~cm$^{-1}$ (particularly for the \cis-isomer), will possess larger errors. Transitions involving these states will have very weak intensities so their inclusion is not expected to significantly impact the computed spectra. However, we would recommend some caution when using the line lists for transitions above 4000~cm$^{-1}$.

\section{Conclusions}
\label{sec:conc}

The rotation-vibration spectrum of diphosphene has been investigated using accurate, first-principles methodologies. High-level \ai\ theory was used to construct new six-dimensional PESs and DMSs which were then utilized in variational nuclear motion computations. Line lists were generated for the \cis- and \trans-isomers of P$_2$H$_2$ which considered transitions between states with energies up to $8000$~cm$^{-1}$ (above the ZPE of each isomer) and total angular momentum $J\leq25$. Comparisons of the fundamental frequencies, their relative intensities, and the ground-state electric dipole moment of \cis-\hpph\ are consistent with previous calculations~\cite{09LuSiEv.P2H2}, giving us confidence in the validity of the presented work. There is, however, a pressing need for quantitative spectral data on the diphosphene molecule as this would enable future benchmarking and the opportunity to improve our theoretical spectroscopic model through empirical refinement of the PESs. 

Regarding future astronomical detection of P$_2$H$_2$, e.g., in the interstellar medium (ISM), the rotational spectrum of \cis-\hpph\ is dipole allowed and appears strong enough at colder temperatures, e.g., at $T=200$~K the intensities of the rotational band are of the order $10^{-22}$~cm/molecule. That said, the presented line list may not be accurate enough in the microwave region and experimentally determined frequencies or more sophisticated \ai\ calculations could be required~\cite{Puzzarini2010}. For the deuterated \trans-P$_2$HD system, which has a non-zero dynamic dipole moment induced by centrifugal distortion, the rotational intensities (of the order of $10^{-24}$~cm/molecule) should be detectable in laboratory studies. 

\section*{Supplementary Material}

See the supplementary material for a list of computed $J=0$ energy levels, \ai\ data points, and the expansion parameters and Fortran routines to construct the PESs and DMSs of diphosphene. 

\begin{acknowledgments}
This work was supported by the STFC Projects No. ST/M001334/1 and ST/R000476/1, and by the COST action MOLIM (CM1405). The authors acknowledge the use of the UCL Legion High Performance Computing Facility (Legion@UCL) and associated support services in the completion of this work, along with the Cambridge Service for Data Driven Discovery (CSD3), part of which is operated by the University of Cambridge Research Computing on behalf of the STFC DiRAC HPC Facility (www.dirac.ac.uk). The DiRAC component of CSD3 was funded by BEIS capital funding via STFC capital grants ST/P002307/1 and ST/R002452/1 and STFC operations grant ST/R00689X/1. DiRAC is part of the National e-Infrastructure.

\end{acknowledgments}

%

\end{document}